\documentclass[preprint,12pt,authoryear]{elsarticle}
\usepackage{graphicx}
\usepackage{amsmath}
\usepackage{amssymb}
\usepackage{setspace}
\usepackage{fullpage}
\journal{Icarus}

\begin{document}
\begin{frontmatter}

\title{Accretion of dust by chondrules in a MHD-turbulent solar nebula}

\author[]{Augusto Carballido}
\ead{augusto@astroscu.unam.mx}
\address{Instituto de Astronom\'{\i}a, Universidad Nacional Aut\'{o}noma de M\'{e}xico,A. P. 70-264, Cd. Universitaria, M\'{e}xico D. F. 04510, Mexico}

\begin{abstract}
Numerical magnetohydrodynamic (MHD) simulations of a turbulent solar nebula are used to study the growth of dust mantles swept up by chondrules. A small neighborhood of the solar nebula is represented by an orbiting patch of gas at a radius of 3 AU, and includes vertical stratification of the gas density. The differential rotation of the nebular gas is replaced by a shear flow. Turbulence is driven by destabilization of the flow as a result of the magnetorotational instability (MRI), whereby magnetic field lines anchored to the gas are continuously stretched by the shearing motion. A passive contaminant mimics small dust grains that are aerodynamically well coupled to the gas, and chondrules are modeled by Lagrangian particles that interact with the gas through drag. Whenever a chondrule enters a region permeated by dust, its radius grows at a rate that depends on the local dust density and the relative velocity between itself and the dust. The local dust abundance decreases accordingly. Compaction and fragmentation of dust aggregates are not included. Different chondrule volume densities $\rho_{c}$ lead to varying depletion and rimmed-chondrule size growth times. Most of the dust sweep--up occurs within $\sim$ 1 gas scale height of the nebula midplane. Chondrules can reach their asymptotic radius in 10 to 800 years, although short growth times due to very high $\rho_{c}$ may not be altogether realistic. If the sticking effiency $Q$ of dust to chondrules depends on their relative speed $\delta v$, such that $Q<10^{-2}$ whenever $\delta v>v_{\rm{stick}}\approx$ 34 cm/s (with $v_{\rm{stick}}$ a critical sticking velocity), then longer growth times result due to the prevalence of high MRI--turbulent relative velocities. The vertical variation of nebula turbulent intensity results in a moderate dependence of mean rimmed-chondrule size with nebula height, and in a $\sim$20\% dispersion in radius values at every height bin. The technique used here could be combined with Monte Carlo (MC) methods that include the physics of dust compaction, in a self-consistent MHD-MC model of dust rim growth around chondrules in the solar nebula.  

\end{abstract}


\end{frontmatter}

\section{Introduction}
The origin of fine-grained dust rims surrounding chondrules in carbonaceous chondrites is still debated. Analyses of CM chondrite thin sections by optical and scanning electron microscopy (Metzler et al. 1992), as well as detailed theoretical models (Morfill et al. 1998, henceforth MDT; Cuzzi 2004; Ormel et al. 2008, henceforth OCT) suggest that these dust mantles could very well have formed in the gaseous environment of the primitive solar nebula, as a result of accretion processes. Nevertheless, advocates of a non-nebular origin of these rims favor their formation on the CM parent asteroids, through the impact and compaction of matrix material around chondrules (Trigo-Rodr\'{\i}guez et al. 2006), followed by aqueous alteration. This interpretation arises mostly because the porosity of observed dust mantles that envelop chondrules is lower than what, presumably, could be produced by agglomeration of dust in the solar nebula. The non-nebular scenario still requires further quantitative analysis. For example, numerical techniques similar to those employed in the study of collisions between Kuiper belt objects (Leinhardt and Stewart 2009) could be used to measure shock stresses in asteroidal collisions.

Theoretical modeling of the formation of fine-grained dust rims assumes that two populations, chondrules and dust, both free floating in the gas of the primitive solar nebula, come into contact. MDT put forward the notion that sweep-up of dust by chondrules in a confined volume of the nebula, prior to the formation of meteorite parent bodies, reproduces the near-linear relation between chondrule radius and rim thickness (e.g. Metzler et al. 1992, Paque and Cuzzi 1997). MDT do not make specific assumptions about the nature of the gas flow. Cuzzi (2004) goes further and incorporates the role of turbulence in the sweep-up process, showing that the initial relative abundance of chondrules is an important parameter that determines both the rate at which dust is depleted locally, and the rate of growth of rimmed particles.

The analytical model of Cuzzi (2004) predicts that chondrule-sized particles will acquire their observed rim volumes in $\sim$10$^{2}$--10$^{3}$ years, subject to the values of the dimensionless turbulent viscosity parameter $\alpha$ (Shakura and Sunyaev 1973, Balbus and Hawley 1998), which is a measure of the stresses in an accretion disk. The precise value of $\alpha$ in protoplanetary disks (including the solar nebula) is uncertain, and authors often consider values ranging from $\sim 10^{-5}$ to 0.1. For disks around T Tauri stars, $\alpha \sim 10^{-4}$--$10^{-2}$ has been inferred (Hartmann et al. 1998). Assuming high values of $\alpha$, the Cuzzi (2004) model yields chondrule rimming times as low as a few years.

The effect of parametrized turbulence has been included by OCT in a numerical scheme to study inter-chondrule sticking via dust rims. Their Monte Carlo algorithm contains a detailed treatment of the collisions among porous dust grain aggregates, and their results evidently show a difference in the evolution of rimmed-chondrule size for two values of $\alpha$, $10^{-6}$ and $10^{-4}$. The lower $\alpha$ value allows for rimmed-particle radii growth by a factor of 7, compared to a factor of $\sim$ 2 for $\alpha=10^{-4}$, before the effects of dust compaction set in. This occurs in $\sim 10^{3}$ years in the former case, and in $\sim 10^{2}$ years in the latter case. The final relation between chondrule and rim masses that OCT obtain is, once again, nearly linear.

A key ingredient in the above calculations is the collision velocity between solid particles of different sizes. In a protoplanetary disk, the following sources of relative velocities may operate: 1) thermally-induced Brownian motion, important for micron-sized dust grains; 2) drift towards the central star and difference in azimuthal (orbital) velocities (the latter also known as ``transverse'' velocity), both due to a radial pressure gradient in the disk gas (e.g. Weidenschilling 1977); 3) settling towards the disk midplane, as a result of the vertical component of the star's gravity; and 4) turbulence. It was only recently that accurate analytical expressions for turbulent relative velocities were obtained (Ormel and Cuzzi 2007). These formulae show that turbulence can dominate mutual velocities for chondrule- and smaller-size particles in the solar nebula at 3 AU, over systematic radial drift, provided $\alpha \gtrsim 10^{-5}$ (OCT, their Figure 1).

Our understanding of the evolution of protoplanetary disks is significantly improved when their vertical structure is taken into account. For example, numerical simulations show that the gas velocity dispersion (Fromang and Papaloizou 2006), viscous stress (Turner et al. 2007), and accretion rate (Turner et al. 2010) depend on height $z$ above the disk midplane. Furthermore, this vertical dependence bears directly on the dynamical behavior of solid bodies. For instance, the increase of gas velocities at high $z$ (where gas densities are lower) produce higher particle relative velocities.

While it is recognized that protoplanetary disks must be in a turbulent state, there is no general consensus regarding the mechanism responsible for the production of turbulence. However, workers in the field acknowledge that a magnetic field, coupled to the differential rotation of the disk gas, could be the crucial driver of disk turbulence. The outcome of the ensuing magnetorotational instability (or MRI; Balbus and Hawley 1998) is self-sustained turbulence that supplies the necessary viscous stresses for the disk angular momentum to be transported outwards. Hydrodynamic and magnetic stresses (the latter greater than the former by a factor of a few) both contribute to the turbulent viscosity in a disk.

There is a caveat associated with the saturation of the MRI in the solar nebula: gas ionization may be too low, and electrical resistivity too high, in dense regions of the nebula to allow coupling of the gas to the magnetic field. These regions remain nearly laminar and are sandwiched between conducting layers where turbulence does develop, owing to ionization by cosmic rays and stellar X-rays that are absorbed before they reach the midplane. Furthermore, recombination of free electrons on micron-sized dust grains is efficient and reduces the ionization fraction below that necessary to sustain turbulence (Turner et al. 2010). Dust abundance is therefore important in establishing levels of turbulent activity.

Notwithstanding the significance of gas resistivity, in this study it is set to zero to focus on the growth of rimmed, chondrule-sized solid particles as they accrete dust in a local neighborhood of the solar nebula, in which all of the gas flow is turbulent due to the MRI and the turbulent intensity varies with time and position. The ideal magnetohydrodynamic (MHD) numerical set-up used allows us not only to determine the final size distribution of dust-mantled objects, but in the process it performs a self-consistent calculation of the dust abundance as a function of time and spatial location, with turbulence and vertical settling as the sources of relative velocities between dust and chondrules.

Section 2 presents a semi-analytical argument regarding the importance of turbulence in dust-chondrule dynamics. Section 3 describes the numerical technique used to model the solar nebula and the solids, including the interaction between chondrules and dust. This Section also lists the simulations that were carried out. Section 4 presents the results of the MHD calculations, and a discusion follows in Section 5. A summary and conclusions are given in Section 6.

\section{Relative velocities between a chondrule and a dust grain}

For a minimum-mass solar nebula (MMSN) model, one can compute the magnitudes of the particle relative velocities as the aerodynamic coupling of the solids to the gas changes, due to the decreasing gas density with height. Figure 1 plots the magnitudes of relative velocities between a 1-mm particle (a typical chondrule size), and a 1-$\mu$m grain, at 3 AU, due to turbulence (\textit{black solid curve}), vertical settling (\textit{dark gray solid curve}) and radial drift (\textit{light gray solid curve}) as a function of $z$ (normalized by the disk scale height $H$). The turbulent curve was obtained from the closed-form analytical formulae of Ormel and Cuzzi (2007):

\begin{equation}\label{eq:oc1}
\left(\frac{\Delta v_{I}}{v_{g}}\right)^{2}  = \frac{St_{i}-St_{j}}{St_{i}+St_{j}} \left(\frac{St_{i}^{2}}{St_{ij}^{*}+St_{i}} -\frac{St_{i}^{2}}{1+St_{i}} - \frac{St_{j}^{2}}{St_{ij}^{*}+St_{j}} +\frac{St_{j}^{2}}{1+St_{j}}\right)
\end{equation}

\begin{eqnarray}\label{eq:oc2}
\lefteqn{ \left(\frac{\Delta v_{II}}{v_{g}}\right)^{2}  = 2\left(St_{ij}^{*}-Re^{-1/2}\right)+\frac{St_{i}^{2}}{St_{ij}^{*}+St_{i}} }\nonumber\\ 
& & -\frac{St_{i}^{2}}{St_{i}+Re^{-1/2}}+\frac{St_{j}^{2}}{St_{ij}^{*}+St_{j}}-\frac{St_{j}^{2}}{St_{j}+Re^{-1/2}}
\end{eqnarray}

\begin{equation}\label{eq:oc3}
\Delta v_{\textrm{turb}}=\sqrt{\left(\Delta v_{I}\right)^{2}+\left(\Delta v_{II}\right)^{2}}
\end{equation}

In these expressions, $St_{i}\equiv \Omega_{\mathrm{eddy}} t_{\mathrm{s},i}$ is the particle Stokes number of particle $i$, which measures the degree of aerodynamic coupling of particles to the gas, with $\Omega_{\mathrm{eddy}}$ the frequency of the largest turbulent eddies [here assumed to be equal to the disk Keplerian frequency, to a good approximation (Carballido et al. 2010)], and $t_{\mathrm{s},i}$ is the particle stopping time (i.e., the time in which a particle equilibrates with the gas when moving relative to it), which for a particle radius $r$ smaller than the mean free path $\lambda$ of the gas molecules is given by  $t_{\mathrm{s}}=r \rho_{\mathrm{p}}/c_{\mathrm{s}}\rho_{\mathrm{g}}$, with $\rho_{\mathrm{p}}$ the particle solid density, $c_{\mathrm{s}}$ the gas sound speed, and
$\rho_{\mathrm{g}}$ the gas density. Furthermore, $\Delta v_{I}$ and $\Delta v_{II}$ are relative velocities induced by eddies that decay in a time longer (class I)  or shorter (class II) than $t_{\mathrm{s}}$; $St_{ij}^{*}$ is a quantity that defines the boundary between classes I and II [obtained as in Section 3 of Ormel \& Cuzzi (2007)]; $v_{\mathrm{g}}$ is the turbulent gas velocity; and $Re=\nu_{\mathrm{turb}}/\nu_{\mathrm{m}}$ is the flow Reynolds number (with $\nu_{\mathrm{turb}}$ and $\nu_{\mathrm{m}}$ the turbulent and molecular viscosities, respectively).

For the purpose of Fig. 1, the Stokes numbers $St_{i}$ and $St_{j}$ are chosen to correspond to particle radii $r$=1 mm and 1 $\mu$m, respectively, with $\rho_{\mathrm{p}}$=3 g cm$^{-3}$. The gas density is assumed to decrease with height as in Eq. (8) below. Note that $t_{\mathrm{s}}$ (and hence $St$) increases with height as the gas density decreases. The Reynolds number is $Re=10^{9}$, with $\nu_{\mathrm{turb}}=\alpha c_{s} H$, $\nu_{\mathrm{m}}=\lambda c_{s}$, and $\lambda$=16.7 cm (Armitage 2010). The vale of $\alpha$ is a vertical profile (\textit{dotted curve} in Fig. 1) obtained by adding together radial and azimuthal averages of the Reynolds and Maxwell stresses from a MHD simulation (Section 3), normalizing the sum by the initial midplane gas pressure, and furter averaging over time. The turbulent gas velocity $v_{\mathrm{g}}$ in Eqs. (\ref{eq:oc1}) and (\ref{eq:oc2}) is also obtained from the MHD run, and is averaged in a similar way.

Two features are evident from Fig. 1: first, the contribution of radial drift to relative velocities is minor, and the main sources are vertical settling and turbulence. Second, turbulent activity, as described by the value of $\alpha$, depends considerably on distance from the midplane. It may appear as if there exist odd anticorrelations between the turbulent relative velocity and $\alpha$ at certain heights (for example, around $|z|=2.4H$); however, the fluctuations in both curves result from the intrinsic variability of MRI turbulence, and in Fig. 1 the time average for $\alpha$ was performed over a different time interval than for $v_{\mathrm{g}}$, so these fluctuations are uncorrelated. The overall trend is for $\Delta v$ to increase with increasing $\alpha$.

\begin{figure*}\label{fig:fig1}
\begin{center}
\includegraphics[width=0.95\textwidth]{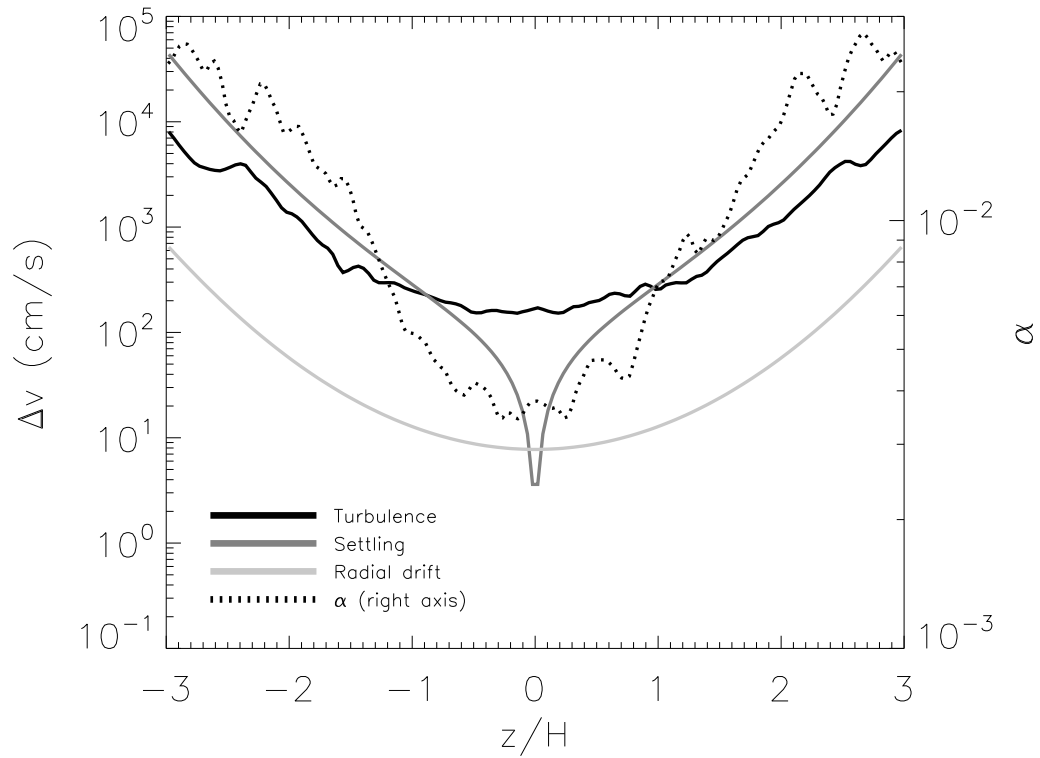}
\end{center}
\caption{Relative velocities $\Delta v$ between a chondrule-size, 1-mm solid particle and a 1-$\mu$m dust grain, due to three different processes, as a function of solar nebula height (in units of nebula scale height $H$). The effects of turbulence (\textit{black solid curve}), vertical settling (\textit{dark gray solid curve}) and inward radial drift (\textit{light gray solid curve}) are shown. The turbulent parameter $\alpha$ (\textit{dotted curve}) also shows a $z$ dependence.  }
\end{figure*}

Numerical simulations of MRI turbulence, in which solid particles are modeled with a Lagrangian approach, show that relative velocities between chondrules and dust can be as high as $\sim$ 10$^{4}$ cm/s (Carballido et al. 2008), higher than what is shown by the semi--analytical black curve in Fig. 1. The intrinsic strength of the velocities produced by the MRI reinforces the main point of Fig. 1, that is, turbulence is a major contributor to dust--chondrule dynamics.

\section{Numerical method}
The model of chondrule-dust compound growth consists of two components: the gaseous (disk) phase and the solid (chondrule, dust and chondrule-dust) phase. Throughout this work, the term particle will be used to denote either a rimless or a rimmed chondrule. 

\subsection{Solar nebula}
The disk is modeled by the shearing box approximation of Hawley et al. (1995): a rectangular Cartesian coordinate system represents a local neighborhood inside the disk, located at an arbitrary orbital radius $R_{0}$ from the central star. The dimensions of this neighborhood are assumed to be much smaller than $R_{0}$. The $x$, $y$ and $z$ directions in the box represent the radial, azimuthal and vertical directions of the disk, respectively. Both the radial and vertical components of stellar gravity are included, and the differential rotation of the disk gas is replaced by a shear flow along the $y$ direction. In this work, the Eulerian ZEUS-3D code (Stone and Norman 1992a,b) is used to solve the equations of ideal MHD in the shearing box:


\begin{equation}
\frac{\partial \textbf{v}_{\textrm{g}}}{\partial t}+\textbf{v}_{\textrm{g}}\cdot\nabla \textbf{v}_{\textrm{g}}= 
 -\frac{1}{\rho_{\textrm{g}}}\nabla\left(P+\frac{B^{2}}{8\pi}\right)
+\frac{\left(\textbf{B}\cdot\nabla\right)\textbf{B}}{4\pi\rho_{\textrm{g}}}-2\mathbf{\Omega}\times\mathbf{v}_{\textrm{g}}
 +\frac{3}{2}\Omega^{2}x\hat{\textbf{x}}-\Omega^{2}z\hat{\textbf{z}}
\end{equation}

\begin{equation}
\frac{\partial\textbf{B}}{\partial t}=\nabla \times\left(\textbf{v}_{\rm{g}}\times\textbf{B}\right)
\end{equation}

\begin{equation}
\frac{\partial \rho_{\rm{g}}}{\partial t} + \nabla\cdot\left(\rho_{\rm{g}}\mathbf{v}_{\rm{g}}\right)=0
\end{equation}

\begin{equation}
P = \rho_{\rm{g}}c^{2}_{\rm{s}}
\end{equation}

\noindent where $\mathbf{v_{\rm{g}}}$, $\rho_{\rm{g}}$ and $P$ are the gas velocity, density and pressure, respectively, $\mathbf{B}$ is the magnetic field, $\mathbf{\Omega}=(0,0,\Omega)$ is the disk angular frequency, $c_{\rm{s}}$ is the isothermal speed of sound, and $\hat{\mathbf{x}}$ and $\hat{\mathbf{z}}$ are unitary radial and vertical vectors.

 The box setup is that of Fromang and Papaloizou (2006). The gas density depends initially on height as   

\begin{equation}\label{eqn:gasdens}
\rho_{\rm{g}}=\rho_{\textrm{\rm{g},mid}}\exp\left(-z^{2}/2H^{2}\right)
\end{equation}

\noindent where $\rho_{\textrm{\rm{g},mid}}$ is the gas density at the midplane, and the disk scale-height $H=c_{s}/\Omega$.

The box has ``shearing periodic'' radial boundary conditions, meaning that whenever a parcel of fluid exits one of the $x$ box walls, it reappears through the opposite wall with a different azimuthal ($y$) position and velocity given by the shear value. Periodic boundaries are applied along $y$ and $z$, but the vertical component of stellar gravity is only included inside a region away from the top and bottom boundaries, to avoid spurious disturbances throughout the box. 

\subsection{Solids}
\subsubsection{Dust}
Small dust grains, which are aerodynamically well coupled to the gas, are represented by a passive contaminant that follows the gas motions. This passive scalar is defined as a fraction of the gas density. The initial dust density is set to 0.1$\rho_{\rm{g}}$, equivalent to about 30 times the solar abundance of silicate dust, since chondrule formation is believed to have taken place in regions of high dust-to-gas ratios (Krot et al. 2010). The particular value of 10\% is somewhat arbitrary, but is intended to represent this enhanced solar abundance. The solid density of dust is assumed to be 3 g cm$^{-3}$.

\subsubsection{Chondrules}
Both rimless and rimmed chondrules are represented by individual particles that interact with the gas through variable drag forces. The particles are smaller than the mean free path of the gas molecules, and their equation of motion is 

\begin{equation}
\frac{d\textbf{v}_{\rm{p}}}{dt} = -\frac{1}{t_{\rm{s}}}\left(\textbf{v}_{\rm{p}}-\textbf{v}_{\rm{g}}\right)
\end{equation}

\noindent where $\textbf{v}_{\textrm{p}}$ is the particle velocity and $t_{\textrm{s}}$ is the stopping time (with the internal density $\rho_{\textrm{p}}$ set to 3 g cm$^{-3}$).

The initial chondrule sizes are either monodisperse ($r$=300 $\mu$m) or drawn from a lognormal distribution with a mean diameter of 622 $\mu$m and a standard deviation of 453 $\mu$m, as measured in the L3 chondrite ALHA77011 (Rubin and Keil 1984). As pointed out by OCT, measured chondrule size distributions may not necessarily reflect the true initial distributions before chondrules were incorporated into chondrite parent bodies. Yet, the process of turbulent concentration \footnote{This mechanism makes particles with a narrow range of aerodynamic stopping times accumulate in regions of low gas vorticity. See Cuzzi and Weidenschilling (2006).} in the solar nebula leads to nearly lognormal size distributions that are consistent with those of disaggregated chondrules from ordinary chondrites (Cuzzi et al. 2001). 

\subsubsection{Dust-chondrule interaction}
Whenever a chondrule enters a region permeated by dust, it is assumed to accrete a dust mantle, subject to the local dust abundance. The effect of collisions between fractal dust aggregates, and between aggregates and chondrules, is not treated here. Instead, we follow the approach of MDT, in which the radius of a rimmed chondrule grows according to

\begin{equation}
\frac{dr}{dt} = \frac{\rho_{\rm{d}}Q}{4\rho_{\rm{p}}}\delta v 
\end{equation}

\noindent where $\rho_{\rm{d}}$ is the spatial mass density of the dust population, $\delta v\equiv v_{\rm{p}}-v_{\rm{g}}$, and $Q$ is the sticking efficiency between a dust grain and a chondrule (see below). Strictly speaking, the chondrule speed $v_{\rm{p}}$ should be measured relative to the dust, but since dust is firmly tied to the gas, $v_{\rm{g}}$ can be used. The corresponding depletion of dust is given by

\begin{equation}\label{eq:deplete}
\frac{d\rho_{\rm{d}}}{dt} = -\int^{\infty}_{0} \pi r^{2}\,Q\, f(r)\,\delta v \,\rho_{\rm{d}}\, dr
\end{equation}

\noindent where $f(r)$ is the particle size distribution function.

Sticking between dust-rimmed chondrules is not investigated here, but previous experimental (G\"{u}ttler et al. 2010 and references therein) and numerical (Carballido et al. 2010, Zsom et al. 2010) work indicates that approximately equal-sized compact particles will not stick at the relative velocities prevalent in ideal-MHD turbulence. Note, however, that OCT show that there is sticking between dust-rimmed chondrules when $\alpha$ becomes large.

\subsection{Simulations}
The shearing box is located at $R_{0}$=3 AU, and has dimensions $(L_{\rm{x}},L_{\rm{y}},L_{\rm{z}}) = (H,2\pi H,6H)$, with the disk scale-height $H$ at 3 AU equal to 0.186 AU. The box contains 32 grid cells in the $x$ direction, 100 cells in the $y$ direction, and 196 cells in the $z$ direction. The number of particles that can be introduced in the numerical disk model is many orders of magnitude less than the actual number in the solar nebula. For the purpose of sampling a large enough region of the turbulent flow, 1.4 $\times 10^{6}$ individual particles are introduced with zero velocity and spread uniformly across the box at $t_{0}$=52 years. In particular, the input vertical particle coordinates are uniformly distributed between $z$=0 and $\pm2.4H$, and are then randomly initialized. The formal approach would have their initial heights be proportional to the vertical gas density profile, but it would make little difference since the mixing time for such small particles is very short.  


The chondrule abundance is introduced through the ratio $\mathcal{R}$ of chondrule volume density $\rho_{\rm{c}}$ to dust density $\rho_{\rm{d}}$. This ratio is a key quantity that determines the final thickness of the dust rims (OCT). As $\mathcal{R}$ is a free parameter, for the purpose of the present numerical runs it is assigned values ranging from $\sim$ 10$^{-1}$ to $\sim$ 10$^{2}$. Table 1 lists the actual values employed in each run (labeled A, B, C, D and E).


Also listed in Table 1 are the corresponding chondrule \textit{number} densities $n_{c}$ [note that chondrule number densities in regions of the solar nebula where these silicate spherules formed may range from $n_{c}\approx$ 2.5$\times10^{-8}$ cm$^{-3}$ (Ciesla et al. 2004) to $n_{c}\sim10^{-5}$ cm$^{-3}$ (Cuzzi \& Alexander 2006)], along with a description of the initial size distribution and the value of the sticking efficiency $Q$. Four runs (A through D) use a constant efficiency $Q$=0.3, while for the last run (E) the following criterion for sticking is introduced: dust sticks with 100\% efficiency if its relative velocity $\delta v$ with respect to a chondrule is less than a critical sticking velocity $v_{\rm{stick}}\approx$ 34 cm/s, given by Eq. (14) of OCT. If $\delta v > v_{\rm{stick}}$, then $Q$ is set to a small, finite value of $8\times 10^{-3}$. Additionally, in run E one half of the chondrule population is started at the midplane ($z=0$) and the other half at $z=2.3H$. The latter case may represent chondrule injection through the surface layers due, for instance, to an X-wind (Shu et al. 2000; however, see comments in this regard in Section \ref{sec:discussion} below).


The properties of gas and dust used in the simulations (density at the midplane $\rho_{\rm{mid}}$, surface density $\Sigma$ and dust solid density $\rho_{\rm{s}}$) are summarized in Table 2. The time step for the evolution of the solid component is the same as the one for the gas, and is typically $\sim 10^{-4}$ years.

\begin{table}[h]
\caption{SIMULATION FEATURES}
\centering
\begin{tabular}{c c c c c}
\hline\hline
Simulation & $\mathcal{R}$   & $n_{c}$ (cm$^{-3})$ & Initial chondrule size distribution & $Q$\\
\hline
A &   0.1 & $4.37\times 10^{-10}$ & Monodisperse & 0.3 \\
B &   10 & $4.37\times 10^{-8}$ & Monodisperse & 0.3 \\ 
C & 600 & $4.37\times 10^{-6}$& Lognormal & 0.3 \\
D &  15 & $1.09\times 10^{-7}$ & Lognormal & 0.3 \\ 
E &  15 & $1.09\times 10^{-7}$ & Lognormal & 1 if $\delta v<v_{\rm{stick}}$, \\ 
  &       &               &           &    $8\times 10^{-3}$ otherwise \\[1ex]
\hline
\end{tabular}
\end{table}

\begin{table}[h]
\caption{GAS AND SOLID PROPERTIES}
\centering
\begin{tabular}{c c c c}
\hline\hline
Component & $\rho_{\rm{mid}}$ (g cm$^{-3}$) & $\Sigma$ (g cm$^{-2}$) & $\rho_{\rm{s}}$ (g cm$^{-3}$) \\
\hline
Gas        & 6.63$\times10^{-11}$   &  327.2   &  --\\
Dust (free-floating)       & 6.63$\times10^{-12}$   &   46.2   &  3 \\ 
Dust (rims)  &  --                    &   --     &  3 \\

\hline
\end{tabular}
\end{table}

\section{Results}
\subsection{Constant $Q$}\label{sec:constq}

Figure 2 shows the dust abundance in runs C (top row) and D (bottom row) at two different times: the instant when dust is introduced in the gas (left panels) and after $t$=83.2 years have elapsed (right panels). Each panel contains the amount of dust in the $y$--$z$ plane, averaged over the radial direction. 
 The effect of different chondrule abundances can clearly be distinguished; a higher chondrule density depletes dust more rapidly (top right panel), and this is more evident close to the midplane, since rimmed chondrules have a higher tendency to settle due to vertical gravity, and therefore spend more time oscillating around $z$=0 (see below). However, with a chondrule abundance that is 40 times smaller (bottom right panel), dust depletion is only starting to become noticeable at the same $t$, once again close to the midplane.

\begin{figure*}\label{fig:fig2}
\begin{center}
\includegraphics[width=0.95\textwidth]{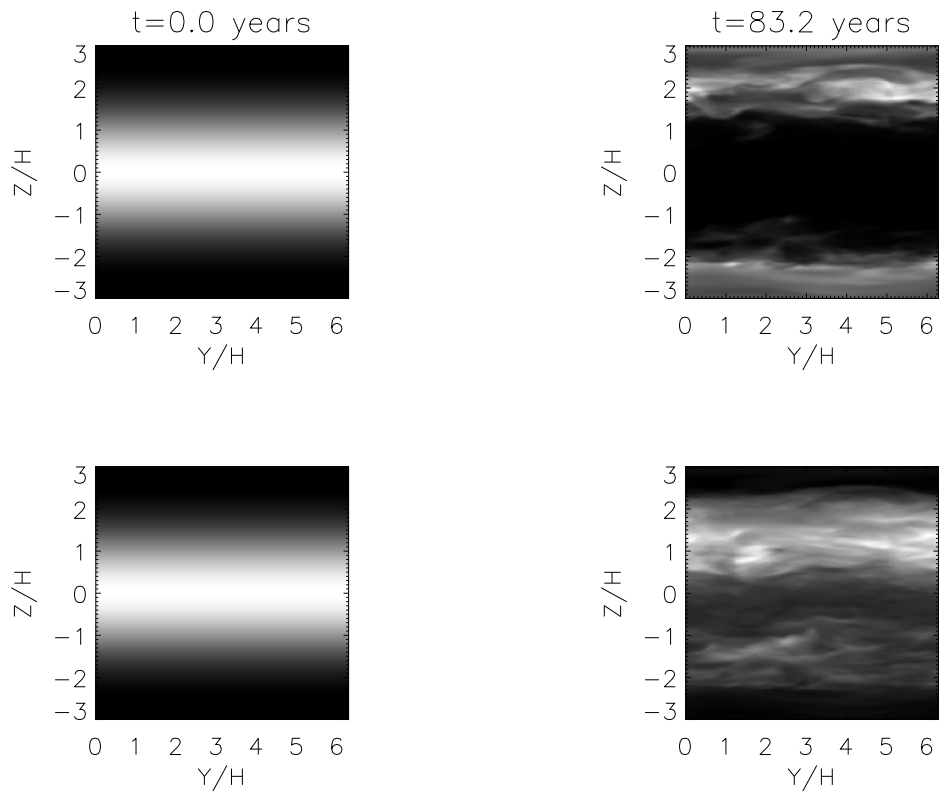}
\end{center}
\caption{Dust density in runs C and D (top and bottom panels, respectively), at the instant when the dust is introduced in the gas (left column) and 83.2 years later (right column). The panels show the radially-averaged dust abundance in the $y$--$z$ plane, normalized by the gas density $\rho_{\rm{g}}$. }
\end{figure*}

The reduction of dust density over time varies significantly if the chondrule abundance differs by a factor of 250, along with a distinct initial chondrule size distribution, as in runs A and D. This is apparent from a vertical profile of $\rho_{\rm{d}}/\rho_{\rm{g}}$, shown in Fig. 3. The curves are $x$--$y$ averages of the dust density. Runs A and D are represented by the \textit{solid} and \textit{dot-dashed} curves, respectively. Black and gray curves correspond to two different snapshots in the evolution of the sytem, $t$=83.2 years (as in Fig. 2) and $t$=306.8 years, respectively. Dust density in run A has decreased by a factor of $\sim$10 over a 224-year period (mostly around the midplane), whereas in run D it has been reduced by up to five orders of magnitude. The chondrule abundance in run C (\textit{dashed} curve) is so high that after 83 years the dust close to the midplane is practically depleted.  

\begin{figure*}\label{fig:fig3}
\begin{center}
\includegraphics[width=0.95\textwidth]{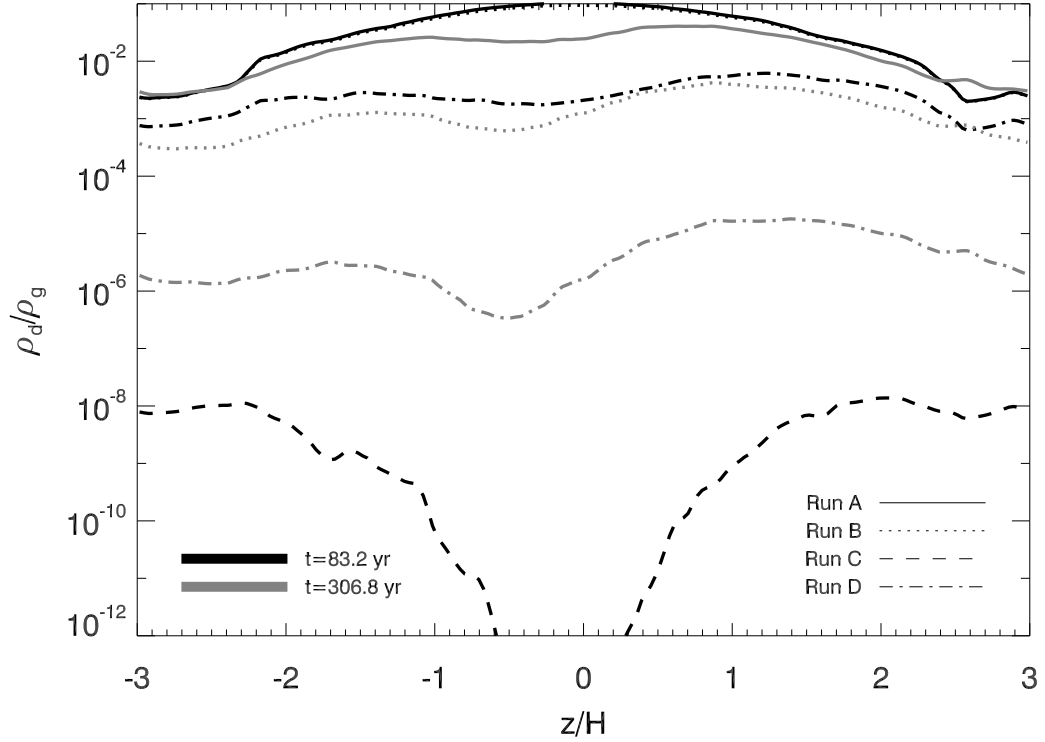}
\end{center}
\caption{Vertical profiles of dust density $\rho_{\rm{d}}$ normalized by the gas density $\rho_{\rm{g}}$. Black curves correspond to profiles 83.2 years after solids are introduced in the turbulent flow, and gray curves represent profiles at $t$=306.8 years. The results for run A (\textit{solid curves}), run B (\textit{dotted curves}), run C (\textit{dashed curve}) and run D (\textit{dot-dashed curves}) are shown. The density profile of run C after 306.8 years is below the scale of the vertical axis, and is not shown here.}
\end{figure*}

Depletion times are more clearly seen in Fig. 4, where the history of volume-averaged $\rho_{\rm{d}}/\rho_{\rm{d,0}}$, with $\rho_{\rm{d,0}}$ the initial dust density, is shown for all runs. Note that in run C, $\rho_{\rm{d}}$ has fallen to 10\% of its initial value in $t=$5 years. That same fall-off occurs for runs A, B and D after 473, 244 and 68 years, respectively. There exists an analytical expression to estimate the time $t_{\rm{col}}$ for a dust grain to collide with a chondrule [OCT, their Eq. (27)], which for the parameters used in the MHD simulations gives $t_{\rm{col}} \approx 1.4$ yr (run A) and $\sim 10^{-5}$ yr (run C). The discrepancy between these values of $t_{\rm{col}}$ and the depletion times inferred from Fig. 4 arises from the spatial average of $\rho_{\mathrm{d}}/\rho_{\mathrm{d},0}$, which is taken over the whole vertical extent of the box, while most of the dust-sweeping process takes place close to the midplane (Fig. 2). 

\begin{figure*}\label{fig:fig4}
\begin{center}
\includegraphics[width=0.95\textwidth]{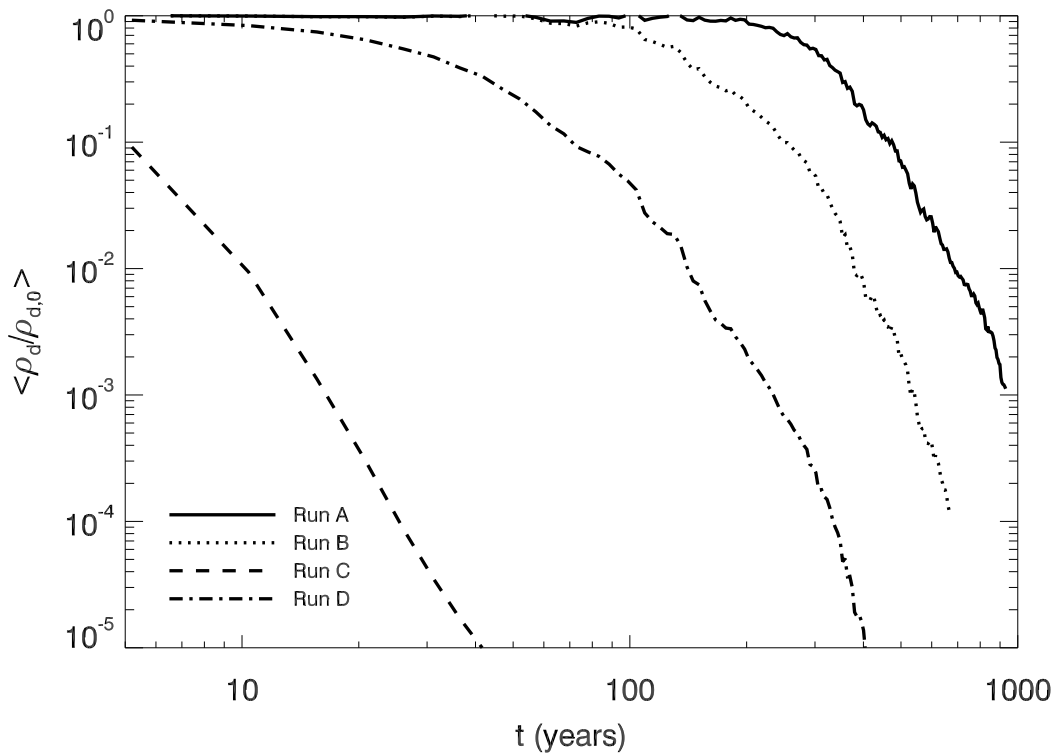}
\end{center}
\caption{Time evolution of the volume-averaged dust density, normalized by its initial value, for constant-$Q$ runs.}
\end{figure*}

Growth of particle radii proceeds at different rates in our simulations, and in all cases an asymptotic size is reached. This size depends strongly on chondrule abundance, as shown by Cuzzi (2004). Furthermore, the final particle size determines the disk height to which particles settle. Figure 5 contains data on particle size and height as a function of time. The black curves describe the behavior of the mass-averaged particle radius for each run, measured on the left vertical axis, while the gray curves depict the evolution of the corresponding root-mean-square particle height, on the right axis. The decrease of the RMS particle height is a consequence of the initially assigned vertical positions. The low chondrule concentration in run A leads to a large rimmed-chondrule radius of almost 1 cm, or more than 30 times the original core size (\textit{black solid curve}). To the author's knowledge, such large rimmed particles have not been reported. Chondrules in this run attain such large sizes at relatively late times, having more dust available to sweep up than in the cases with higher chondrule concentration. During the period of maximum growth, between 0 and $\sim$ 200 years, these particles experience considerable vertical settling from their initial RMS height (\textit{gray solid curve}), and oscillate thereafter around $z \approx 0.32H\approx$ 0.059 AU above and below the midplane.

\begin{figure*}\label{fig:fig5}
\begin{center}
\includegraphics[width=0.95\textwidth]{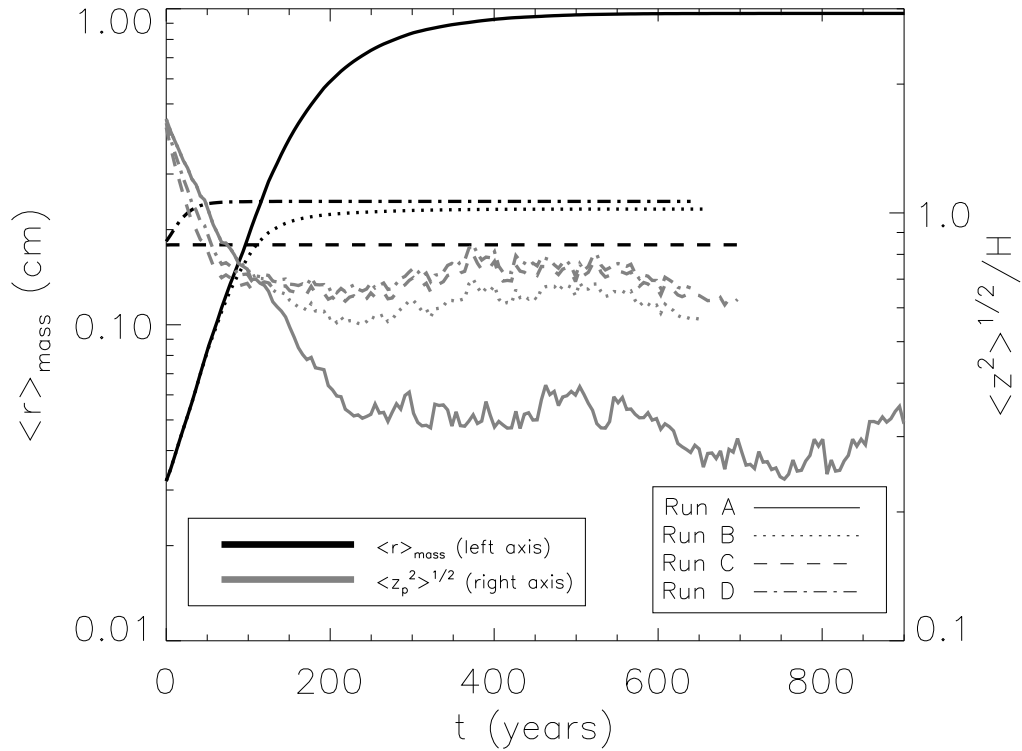}
\end{center}
\caption{Mass-averaged particle radius (\textit{black curves}, measured on left axis) and root-mean-square particle height (\textit{gray curves}, measured on right axis) as a function of time. Data for runs with constant $Q$ are shown. }
\end{figure*}

In contrast, the particles in run C (\textit{dashed curves}) exhibit a radius increase of only 0.1\%, imperceptible on the scale of Fig. 5. They undergo a $\sim$100-year period of settling and remain at about $|z|\approx 0.68H$, or 0.126 AU. This is a similar final height as that of run D particles (\textit{dot-dashed curves}, $|z|\approx 0.72H\approx 0.133$ AU), which acquire a radius of roughly 0.25 cm, or 34\% larger than their chondrule cores, in $\sim$50 years.  

It is interesting to make comparisons among the final particle size distributions that result from these simulations. Figure 6 shows probability distribution functions of particle radius at the end of each run. The histograms consist of 100 radius bins each (with equidistant bin spacing), and each bin contains $N_{\rm{b}}$ particles out of $N_{\rm{tot}}=1.4\times 10^{6}$. Not surprisingly, the low chondrule abundance run (\textit{black curve}) exhibits a distribution that peaks towards the high end of the size range. Both runs A and B produce a size distribution even though they begin with a monodispersion (the \textit{black, dashed} vertical line represents the distribution in runs A and B at $t$=0 years). The initial lognormal distributions in runs C and D are indistinguishable from the curve that corresponds to run C (which is the run that exhibits the least growth). In run D (\textit{lightest-gray curve}), particle sizes remain lognormal-distributed, with a 37\% increase in modal radius. 


\begin{figure*}\label{fig:fig6}
\begin{center}
\includegraphics[width=0.95\textwidth]{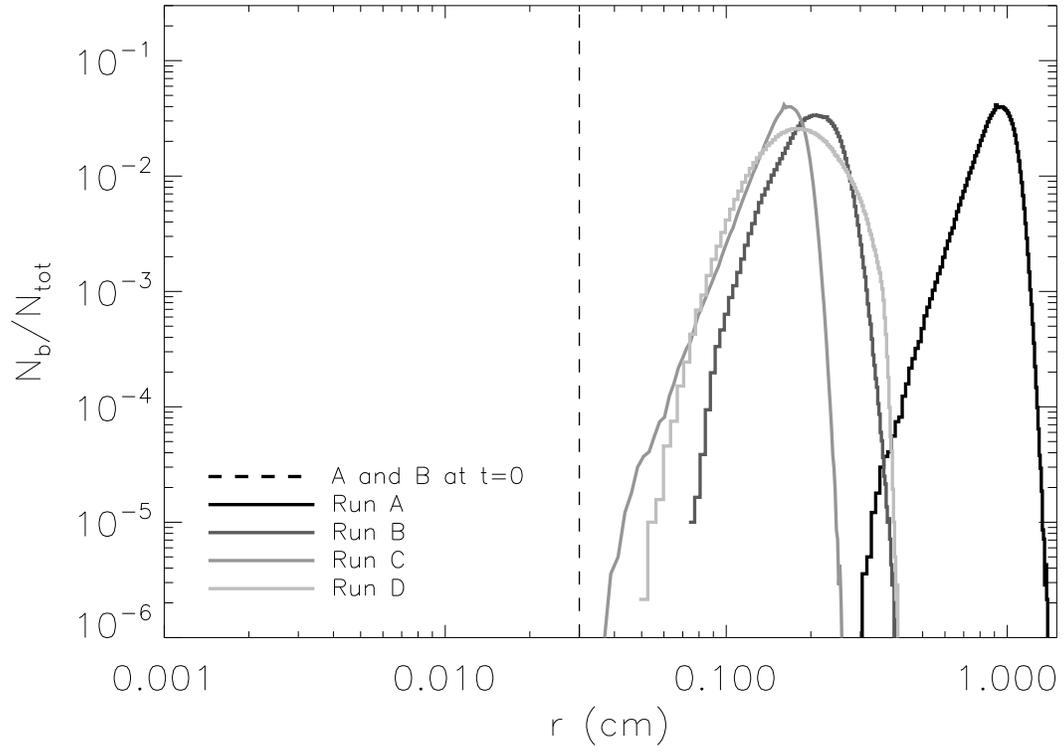}
\end{center}
\caption{Probability distribution functions of rimmed-chondrule radii at the end of each constant-$Q$ simulation (\textit{solid curves}). The \textit{black, dashed line} stands for the initial distribution in runs A and B at $t$=0 years. The initial lognormal distributions of runs C and D are indistinguishable from the curve corresponding to run C.} 

\end{figure*}

Runs A--D show a moderate dependence of particle radius with disk height, as shown in Fig. 7. The curves were obtained by binning particle radii at different values of $z$, and averaging over time. The radii values for run A (\textit{black curve}) have been scaled by a factor of 4.5 to make comparison with the other runs easier. In run A, particle radii at the midplane are $\sim$17\% larger than at $\sim 1.3H$, which is the highest location in that run where rimmed chondrules can be found. Likewise, average radii values in run C (the fastest-growing particles) are larger by approximately the same amount at the midplane than in the upper layers. For the case of run D (top curve), the difference in mean radii at these two locations is $\sim$12\%. 

\begin{figure*}\label{fig:fig7}
\begin{center}
\includegraphics[width=0.95\textwidth]{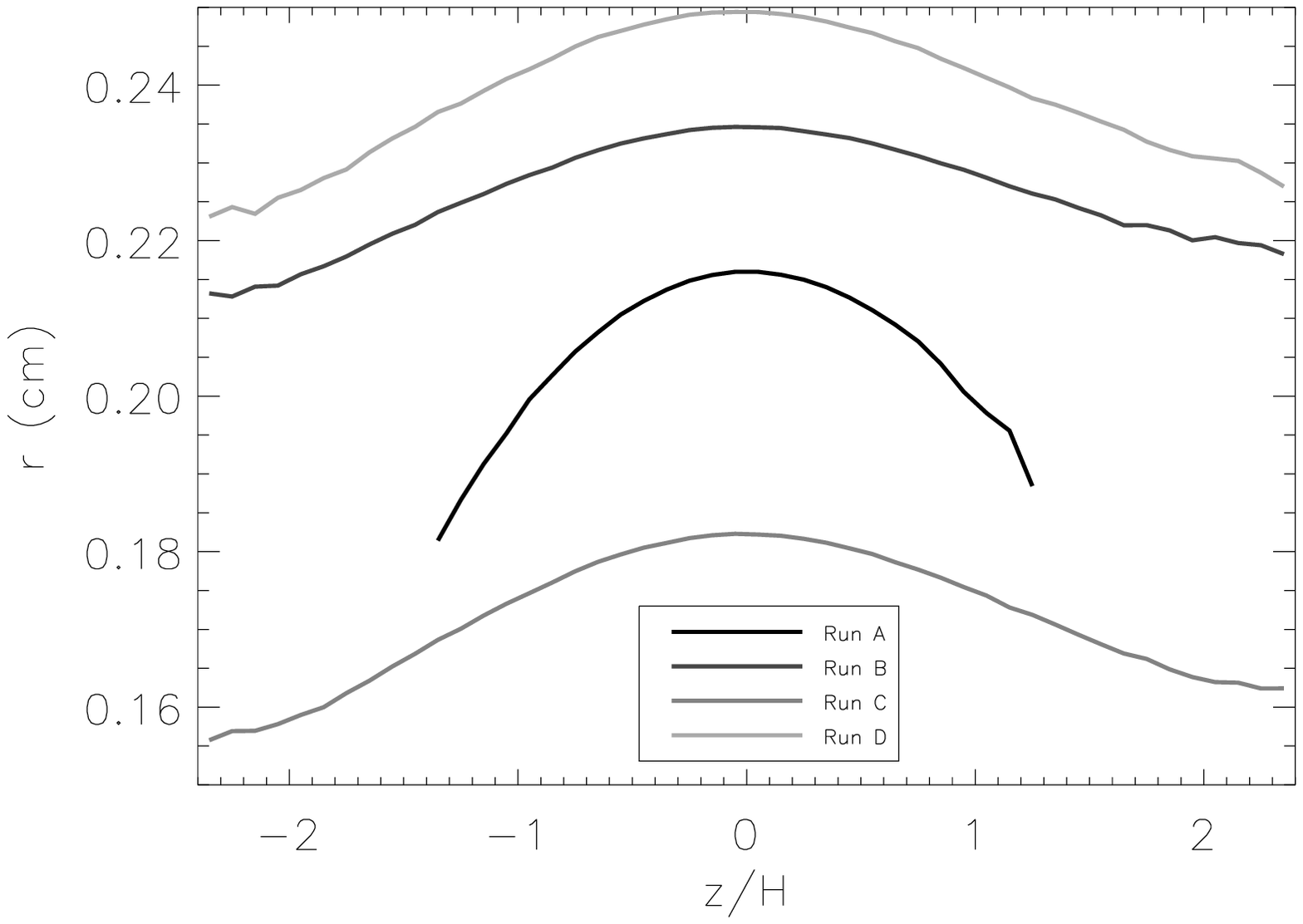}
\end{center}
\caption{Binned particle radii as a function of disk height, averaged over time, for constant-$Q$ runs. Data for run A (\textit{black curve}) have been decreased by a factor of 4.5 for illustrative purposes. Error bars are omitted for clarity, but are typically 20\% of the values shown, a relatively large scatter. }
\end{figure*}

The data in Fig. 7 have a typical scatter of $\sim$20\% (error bars have been omitted for clarity). This means that in run D dust envelopes around chondrules can be as small as 1.8 mm in radius in the upper layers, and as large as 3 mm at the midplane, a 67\% difference. On the other hand, the relatively large dispersion also means that particles can be found whose sizes are larger in the upper layers by 70\%, compared to the lowest regions of the solar nebula.

\subsection{Double-valued $Q$}

The evolution of the mass-averaged particle radius (\textit{black curves}) in run E, for the first 300 years, is shown in Fig. 8. The chondrule population that starts at the upper layer is represented by the dashed curves, whereas the solid curves correspond to the midplane population. The higher dust abundance at the disk midplane results in a systematically greater $\langle r\rangle_{\rm{mass}}$ for the chondrules that are originated at that location, compared to the radius of the upper-layer particles. The mean difference is 1\%. The very low sticking efficiency, which is prevalent due to high relative speeds between chondrules and dust, leads to the slowest radius growth rate of the five runs. These relative speeds can reach up to $\sim$ 10$^{4}$ cm/s, well above the critical value for sticking of 34 cm/s derived from Eq. (14) of OCT.

\begin{figure*}\label{fig:fig8}
\begin{center}
\includegraphics[width=0.95\textwidth]{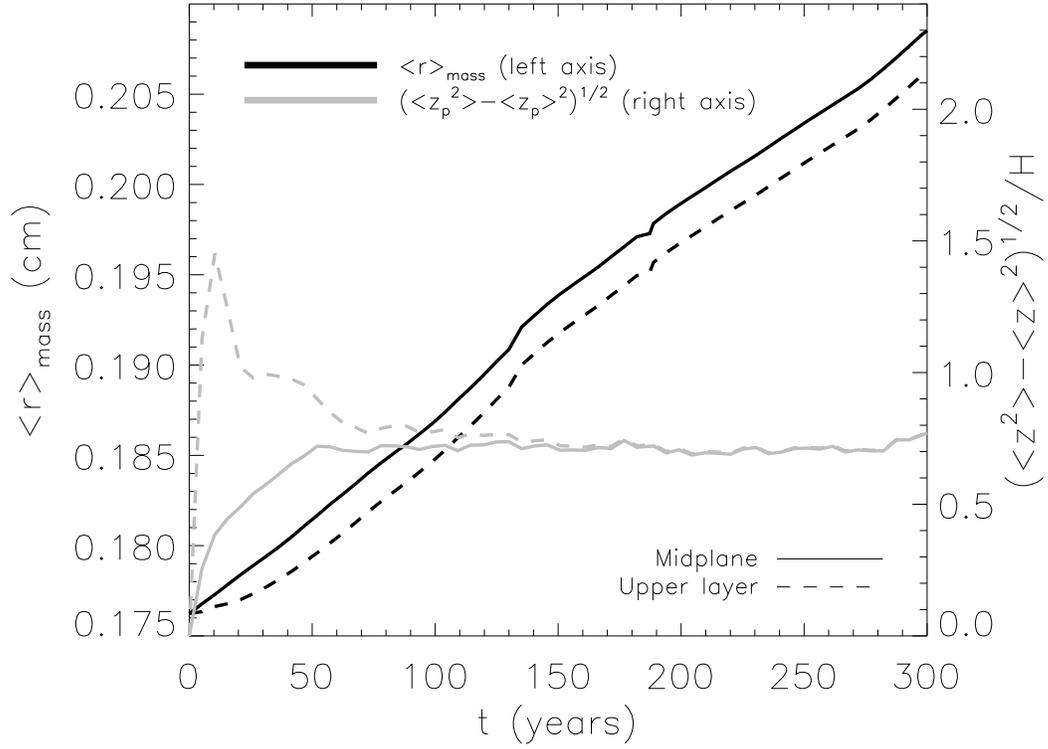}
\end{center}
\caption{Mass-averaged particle radius (\textit{black curves}, measured on left axis) and root-mean-square particle height (\textit{gray curves}, measured on right axis) as a function of time, for run E. }
\end{figure*}

The root-mean-square particle heights in Fig. 8 (\textit{gray curves}) for both populations converge at 0.7$H$. The early sharp rise of the dashed gray curve reflects the fast vertical spreading (i.e., increase in chondrule sheet thickness) of the upper-layer chondrules due to more vigorous turbulent activity in those regions. 

Dust density, measured either through its vertical profile or through the time evolution of its volume-averaged value, does not show significant variation during the course of run E, indicating that dust deplenishment occurs on time scales longer than 300 years for the parameters used, in contrast to the results of Figs. 3 and 4. Of course, this can change for higher chondrule abundances.

\section{Discusion}\label{sec:discussion}
This is a first attempt at understanding the acquisition of dust mantles by chondrules in a magnetohydrodynamic environment. The numerical solar nebula model used here contains several simplifications, of which perhaps its local character is the most restrictive, because potentially important global dynamics are absent. One example is the meridional flow pattern that transports passive species over radial distances of several tens of AU (Wehrstedt and Gail 2008, Ciesla 2009), with inward motion in the upper layers and outward drift close to the midplane. Ciesla (2009) found that the concentration of 500-$\mu$m particles has a maximum at $\sim$10 AU in the midplane of a protoplanetary disk with a mass accretion rate corresponding to that of a T Tauri system, $\dot{M}\approx 10^{-8} M_{\odot}$ yr$^{-1}$. This maximum is 57\% greater than the abundance at 3 AU, and it comes about as a result of the radial transport of material in a $10^{6}$-year old disk. Although the timescales for dust rimming of chondrules are significantly shorter than global transport of sub-millimeter particles (and likely also shorter than transport of millimeter-sized particles), the end result of solid matter transfer between the inner and outer regions of the solar nebula may alter the chondrule abundance at the orbital radius of interest.

OCT calculate the evolution of dust density as porous aggregates accrete onto chondrules, for their two nominal values of $\alpha$, $10^{-4}$ and $10^{-6}$ (their Fig. 6). In their numerical model, $\rho_{\rm{d}}$ decreases to 0.1\% of its initial value in $\sim$3000 ($\alpha=10^{-6}$) and 1000 ($\alpha=10^{-4}$) years. While such long times may seem at odds with the collision time scales mentioned in Section \ref{sec:constq}, it is important to bear in mind that the low values of $t_{\mathrm{col}}$ originate from the high relative velocities imparted by MRI turbulence, which OCT do not model. In our simulations, with $\alpha$ between 3.5$\times10^{-3}$ and 2$\times10^{-2}$ (Fig. 1) being a function of position and time, long depletion times of $\sim$ 1000 years occur only for very low chondrule abundance, run A (Fig. 4). Run D depletes dust to $10^{-5}\rho_{\rm{d},0}$ in 400 years. Clearly, $\alpha$ plays a significant role in how much dust is available for chondrule rim growth. In fact, rapid growth can be understood by considering the dependence on $\alpha$ of the relative velocity $\delta v$ between particles and gas [a proxy for the relative velocity between particles and dust in Eq. (7)], $\delta v \sim \alpha^{1/2}$ (OCT).  

The importance of $\alpha$ will become more evident in calculations that allow for the development of a disk dead zone, the nearly-laminar region of low viscous stresses mentioned in Section 1 that results from non-ideal MHD effects. The degree of coupling of magnetic fields to the disk gas will determine whether the flow is susceptible to the MRI. Magnetically-active regions, which are ionized by stellar X-rays or interstellar cosmic rays, will be in a turbulent state, whereas dense zones in which recombination processes are fast, such as the disk interior, will be quiescent. However, minimum-mass solar nebula models with time-dependent gas ionization chemistry (e.g. Turner et al. 2007) show that ionized gas can reach the disk interior through turbulent mixing and, under certain conditions, this will allow a weak coupling to magnetic fields and an increase in turbulent intensity. The assumption of a wholly turbulent disk used in this work relies on such results, as well as on the fact that electron fractions as low as $\sim 10^{-13}$ can render the flow unstable to the MRI (Balbus 2009). In any case, a marked vertical gradient of turbulent activity is expected at some point. Variable-Ohmic-resistivity calculations by Turner et al. (2007) yield mean magnetic stresses that are 20 times larger in the upper layers than at the midplane, at an orbital radius of 1 AU. A variation by a factor of 100 is found at 5 AU. The original computations of dead zone evolution in a shearing box (Fleming and Stone 2003) obtained $\alpha \approx 6.3\times 10^{-5}$ at the midplane and $\alpha\approx 2.4\times 10^{-3}$ at heights $|z|$=2$H$. Based on these figures, the work of OCT is likely to describe conditions inside a solar nebula dead zone, which may have a limited lifetime. Increasing turbulent activity that originates at the top and bottom layers of the nebula can augment $\alpha$ close to the midplane by a factor of $\sim$ 37 (Turner et al. 2007).


The well-documented linear relation between dust rim thickness and chondrule radius is evident at the end of run D (Fig. 9, \textit{black circles}) and E (\textit{asterisks} and \textit{squares}, which correspond to miplane and upper-layer chondrules, respectively). Data from disaggregated rimmed chondrules in CV chondrites (Paque \& Cuzzi 1997, \textit{triangles}) are shown for comparison. The straight line has a slope of unity. The scatter in the disaggregation data reflects the range of values that can be measured in real chondrites. The linear trend derived from runs D and E would need to be verified, once again, with the inclusion of shattering processes that may produce a realistic scatter of rim thicknesses. It is worthwhile noting that Trigo-Rodr\'{\i}guez et al. (2006) found only a modest correlation between dust rim thickness and size of the enclosed core in CM chondrites.

\begin{figure*}
\begin{center}
\includegraphics[width=0.95\textwidth]{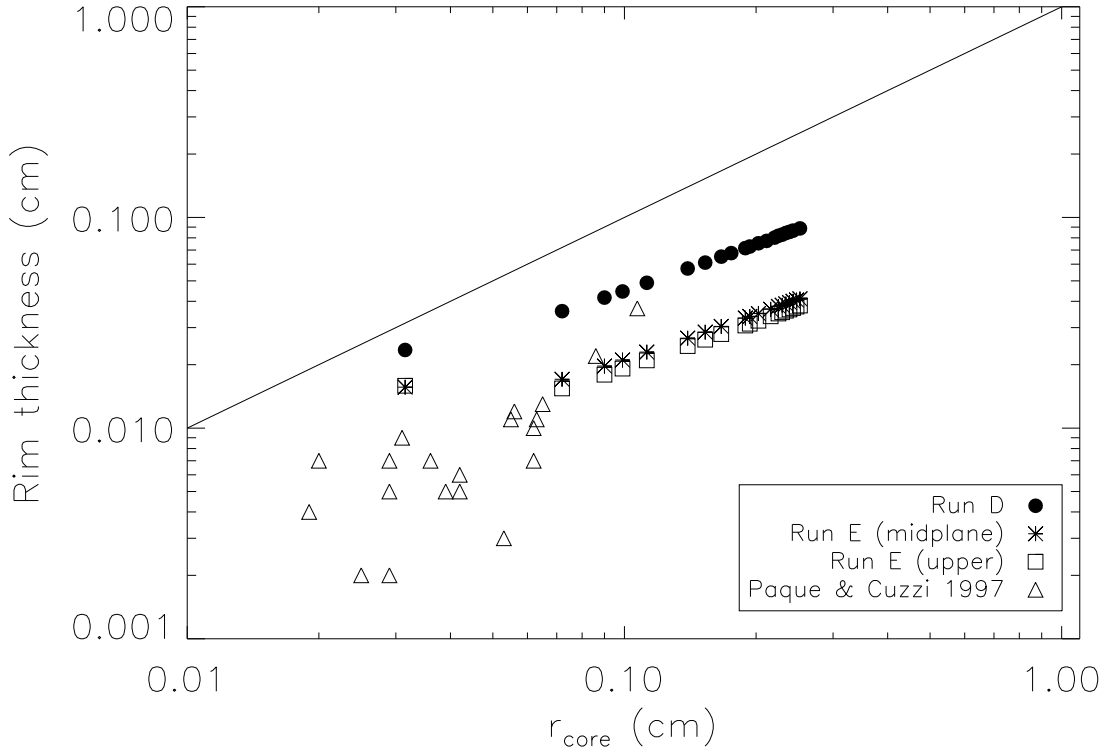}
\end{center}
\caption{Dust rim thickness as a function of chondrule core radius at the end of runs D (\textit{black circles}) and E (\textit{asterisks}, midplane chondrules; \textit{squares}, upper-layer chondrules), and from disaggregation measurements of CV chondrites by Paque \& Cuzzi (2007, \textit{triangles}). The straight line has a slope equal to 1. }
\end{figure*}

The separation of chondrules into two distinct populations in run E, according to their initial vertical position $z$ (0 or 2.3$H$), served the purpose of making a dynamical comparison between chondrules formed in the disk interior and those that could have been dumped in the upper layers after an X-wind carried them from the nebular X region. However, the very formation of chondrules at the X point has been called into question due to dynamical and thermal inconsistencies within the X-wind model (Desch et al. 2010). It is therefore important to bear in mind that a population of dust-rimming chondrules deposited high up in the solar nebula may not come about easily.

\section{Summary, conclusions and future work}

In this work, accretion of dust by chondrule-sized particles is modeled using a MHD representation of a turbulent solar nebula. MHD processes are a viable mechanism to produce and sustain turbulent viscosity in protoplanetary disks, and the dynamical evolution of the ionized gas in the solar nebula must depend crucially on its ability to couple to different magnetic field configurations. Relative velocities between chondrules and micron-sized dust grains can be excited by several sources, and here we focused on the effect imparted by MRI turbulence and vertical settling due to stellar gravity, in a local neighborhood of the nebula. An important parameter in determining the rate of depletion of dust due to sweep-up by chondrules is the latter's volume density. The MHD simulations show that chondrules accrete dust mostly within $\sim$ 1 scale-height ($H$) of the nebula midplane, since the growth of dust rims decreases aerodynamic coupling of particles to the gas, and increases their response to vertical gravity. Time scales to deplete dust to 1\% of its initial abundance vary from 10 to $\sim$ 800 years, roughly the same times in which chondrules reach their mass-averaged, asymptotic radius. If the sticking efficiency of dust to chondrules is below $10^{-2}$, the growth time scale could be $\sim 10^{3}$ years. Low sticking eficiencies (i.e., very little dust rim growth) can be common due to very high, MRI--driven, turbulent relative velocities between chondrules and dust. 

Vertical stratification of the nebular gas density allows to compare the dust rimming process at different nebula heights, and this has been done by initially placing one chondrule population at the nebula midplane and another population at a height of 2.3$H$. The growth rate of dust rims in these two cases is very similar, at least for the first 300 years of evolution, with the midplane population having a mass--averaged radius $\sim$ 1 \% larger than the upper layer population during the course of the numerical run. 

Size distributions of rimmed chondrules obtained from the MHD simulations could be used for comparison with data from actual chondrites. The aerodynamic properties of chondrules suspended in nebular gas are reasonably well understood, and such comparison would provide further insight into the MHD conditions in the primitive solar nebula. In particular, the level of turbulence, and hence the ionization state of the gas, could be constrained. Sweep-up of dust by chondrules is likely to influence the evolution of a laminar dead zone in the solar nebula. Future MHD calculations should incorporate explicit ionization processes (such as those due to stellar X-rays and radionuclide decay) to obtain a more complete, layered flow structure in which chondrule rimming at different locations can be characterized.

The vertical variation of turbulent intensity and dust abundance leads to a modest dependence of rimmed-chondrule size on disk height. Whether this variation could be imprinted and identified on the meteoritic record will require further analysis to account for processes such as fragmentation and compaction of dust structures. Compaction of initially fractal dust aggregates can result from collisions, and compact dust rims can play a role in the sticking between rimmed chondrules. Monte Carlo (MC) methods (OCT, Zsom and Dullemond 2008) provide a robust procedure to investigate coagulation of chondrule compounds, since they treat interactions \textit{between} dust--rimmed chondrules, providing an extension to the present work. These codes take as one of their inputs particle--particle relative velocities $\Delta v$, which could be taken directly from a MHD simulation (e.g., by measuring $\Delta v$ between Lagrangian particles located inside an Eulerian grid cell, as in Carballido et al. 2010). Although a MC code does not resolve the spatial structure of the protoplanetary disk, it contains all the physics of particle--particle collisions. On the other hand, the MHD run provides the spatial structure of the nebula plus position--dependent relative velocities, which can be plugged back into the MC code to provide self--consistency to rimmed chondrule--chondrule collisions. Furthermore, if non--ideal MHD effects are allowed for (e.g., a dead zone), the growth of chondrule compounds in both turbulent and nearly laminar regions could be characterized and related to the time-- and space--dependent ionization state of the gas, which may be useful in chemical studies of chondrule compounds. 


The use of larger shearing boxes now allows to study locations of protoplanetary disks that reach up to 30 scale heights in radial extent (Stone and Gardiner 2010). Thus, in addition to turbulence and vertical settling, the contribution of radial drift to relative velocities between chondrules and dust could be incorporated self-consistently in a MHD-Monte Carlo numerical scheme. The combination of these two techniques is likely to set the nebular origin of dust mantles on a stronger footing.  

\section*{ACKNOWLEDGEMENTS}
I am grateful to Jeffrey Cuzzi for comments and suggestions on an early version of this manuscript. Data from chondrule disaggregation work were kindly provided by Julie Paque and Jeffrey Cuzzi. Valuable remarks by the referees, Chris Ormel and Fred Ciesla, greatly improved the contents and structure of the paper. This work was supported by DGAPA at UNAM.

\end{document}